# Nonlinear dynamics of magnetohydrodynamic flows of heavy fluid over an arbitrary surface in shallow water approximation

**Karelsky K.V.** [1], **Petrosyan A.S.** [1,2], **Tarasevich S.V** [1,3].

[1] Space Research Institute, 84/32 Profsoyuznaya st, 117997 Moscow, Russia.

[2] Moscow Institute of Physics and Technology, 9, Institutskii per., 141700 Dolgoprudny, Moscow Region, Russia

[3] Moscow State University, Department of mathematics and mechanics, 1A, Leninskie gory st, 119234 Moscow, Russia.

E-mails: kkarelsk@iki.rssi.ru (Karelsky K.V.), apetrosy@iki.rssi.ru (Petrosyan A.S.) aprilfire.ru@gmail.com (Tarasevich S.V.)

**Abstract**

The magnetohydrodynamic equations system for heavy fluid over an arbitrary surface in shallow water approximation is studied in the present paper. It is shown that simple wave solutions exist only for underlying surfaces that are slopes of constant inclination. All self-similar discontinuous and continuous solutions are found. The exact explicit solutions of initial discontinuity decay problem over a flat plane and a slope are found. It is shown that the initial discontinuity decay solution is represented by one of five possible wave configurations. For each configuration the necessary and sufficient conditions for its realization are found.

The change of dependent and independent variables transforming the initial equations over a slope to those over a flat plane is found.

**Introduction**

The shallow water magnetohydrodynamic (SMHD) equations are the alternative to solving of full set of magnetohydrodynamic equations for a heavy fluid with a free surface. These equations are derived from the magnetohydrodynamic equations for an incompressible nonviscous fluid layer in the gravity field assuming the pressure is hydrostatic, using the depth averaging, and considering that the fluid layer depth is much smaller than the problem characteristic size. The derived system [1-2] is important in many applications of magnetohydrodynamics to astrophysical and engineering problems. The magnetohydrodynamic shallow water approximation is widely used for the solar tachocline study [1, 3-5], for the description of spread of matter over a neutron-star surface during disc accretion [6-7], for the study of neutron-star atmosphere dynamics [8-9], for the study of extrasolar planets [10], for the optimization of aluminum production process [11-12] and in fusion technologies [13].

Nonlinear dynamics of above flows is described by the full set of magnetohydrodynamic equations for all scales. This system cannot be examined analytically and still is difficult for a computer modeling. Practically, shallow water approximation acts the same fundamental part in the plasma magnetohydrodynamics as the similar approximation acts in the neutral fluid dynamics [14-15]. The latter case is used widely to study the large-scale processes in Earth atmosphere and oceans [16]. In simple case, when neglecting rotation and other external forces, the magnetohydrodynamic shallow water equations over a flat plane are Hamiltonian [17].

The present work is devoted to the study of nonlinear flows of heavy fluid described by the shallow water magnetohydrodynamic equations over a non-flat surface. This new set of equations provides general interests in fluid dynamics. Moreover, the Coriolis force and other external forces in the large scale magnetohydrodynamic models can be naturally represented by using an effective non-flat surface, as it is in the physics of large-scale atmospheric and oceanic flows. These equations serve the basics in developing of multilayer stratified shallow water magnetohydrodynamic models, and in developing of finite volume numerical methods for magnetohydrodynamic shallow water flows subjected to an external force (e.g. Coriolis force or a hydraulic friction).

The hyperbolicity of magnetohydrodynamic shallow water equations leads to the existence of discontinuous solutions as well as the existence of continuous ones. The equations nonlinearity and hyperbolicity can lead to the discontinuous solutions even if the initial conditions are differentiable. In the present paper simple wave solutions for the SMHD equations over a non-flat plane are studied. It is shown these solutions exist only for the class of underlying surfaces that are slopes of constant inclination. Magnetogravity rarefaction wave, magnetogravity shock wave and Alfvenic wave solutions for slopes are found. Characteristics of these waves are parabolas transforming to straight lines in case of flat plane. These particular waves are fundamental for studying of nonlinear wave phenomena over a non-flat surface. The change of dependent and independent variables transforming the SMHD equations on a slope to those on a flat plane is found through analyzing the obtained solutions. It is used to find the exact solution of the initial discontinuity decay problem for SMHD equation system over a slope. We found that the structure of the solution over a slope is the same as for a flat plane. The conditions of each wave configuration realization are exactly match. The initial discontinuity decay solution is represented by one of the following wave configurations: 'two magnetogravity shock waves, two Alfvenic waves', 'magnetogravity shock wave, magnetogravity rarefaction wave turned forward, two Alfveinc waves', 'magnetogravity rarefaction wave turned back, magnetogravity shock wave, two Alfvenic waves', 'two magnetogravity rarefaction waves, two Alfvenic waves', 'two hydrodynamic rarefaction waves and a vacuum region between them'.

In the first section of present article the initial equations of shallow water magnetohydrodynamics over an arbitrary surface are present. In the second section this set of equations is written in Riemann invariant form and it is shown that simple wave solutions exist only for underlying surfaces that are slopes of constant inclination. In the third section particular continuous and discontinuous simple wave solutions are found. In the fourth section the initial discontinuity decay problem solution for flat plane and for slope is found. The main results of present work are outlined in the conclusion.

## 1. Shallow water magnetohydrodynamic equations over an arbitrary surface

In this section we consider the magnetohydrodynamic shallow water model to study the magnetic fluid flows with a free surface in the gravity field over an arbitrary boundary. The magnetohydrodynamic shallow water equations over an arbitrary boundary are obtained from the classical magnetohydrodynamic equations [18] written for the fluid layer with a free surface in the gravity field over an arbitrary boundary. There the Z axis is parallel to the gravity force vector and is opposite in direction to that of gravity force. Assuming the layer depth is small compared to the characteristic size of the studied phenomena and the full pressure (the sum of magnetic and hydrodynamic pressures) is hydrostatic the mentioned system is averaged over a fluid layer depth neglecting the squares of velocity and magnetic field deviations. We denote $\tilde{B}_i = B_i / \sqrt{\rho}$ to simplify the equations (the tilde sign is omitted below) and write this system in one-dimensional case:

$$\frac{\partial h}{\partial t} + \frac{\partial hu_1}{\partial x} = 0 \tag{1.1}$$

$$\frac{\partial hu_1}{\partial t} + \frac{\partial \left(hu_1^2 - hB_1^2 + gh^2/2\right)}{\partial x} = -gh\frac{\partial b}{\partial x} \tag{1.2}$$

$$\frac{\partial hu_2}{\partial t} + \frac{\partial \left(hu_1 u_2 - hB_1 B_2\right)}{\partial x} = 0 \tag{1.3}$$

$$\frac{\partial hB_1}{\partial t} = 0 \tag{1.4}$$

$$\frac{\partial hB_2}{\partial t} + \frac{\partial \left(hB_2 u_1 - hB_1 u_2\right)}{\partial x} = 0 \tag{1.5}$$

$$\frac{\partial hB_1}{\partial x} = 0 \tag{1.6}$$

There $x$ and $t$ are the spatial and temporal coordinates correspondingly, $h(x,t)$ is the fluid depth, $u_1(x,t)$ and $u_2(x,t)$ are fluid velocities along X and Y axes respectively, $B_1(x,t)$ and $B_2(x,t)$ are magnetic field components along X and Y axes respectively and $g$ is the gravitational constant. This system is known as the magnetohydrodynamic shallow water system over an arbitrary surface (see James A. Rossmanith PhD dissertation, University of Washington 2002, chapter 4). Equation (1.6) is the consequence of magnetic field divergence-free equation and is used to set the correct initial data.

In the next section we find the simple wave solutions of this system.

## 2. Riemann waves for SMHD equations over an arbitrary surface

In this section we rewrite the initial equations (1.1-1.6) in the form for Riemann invariants that is more appropriate for further consideration.

It immediately follows from equations (1.4, 1.6) that

$$hB_1 = const \qquad (2.1)$$

Let us rewrite equation (1.1) in the form $\frac{\partial h}{\partial t} = -u_1 \frac{\partial h}{\partial x} - h \frac{\partial u_1}{\partial x}$. Thus time derivatives in the initial equations are transformed to the form

$$\frac{\partial h\alpha}{\partial t} = h\frac{\partial \alpha}{\partial t} + \alpha\left(-h\frac{\partial u_1}{\partial x} - u_1\frac{\partial h}{\partial x}\right) \qquad (2.2)$$

where $\alpha = u_1, u_2, B_1, B_2$.

Using expressions (2.2) for time derivatives in the equations (1.1-1.6) we obtain

$$\partial_t \begin{pmatrix} h \\ u_1 \\ u_2 \\ B_2 \end{pmatrix} + \begin{pmatrix} u_1 & h & 0 & 0 \\ \frac{c_g^2}{h} & u_1 & 0 & 0 \\ 0 & 0 & u_1 & -B_1 \\ 0 & 0 & -B_1 & u_1 \end{pmatrix} \partial_x \begin{pmatrix} h \\ u_1 \\ u_2 \\ B_2 \end{pmatrix} = \begin{pmatrix} 0 \\ -g\frac{\partial b}{\partial x} \\ 0 \\ 0 \end{pmatrix} \qquad (2.3)$$

$$hB_1 = const \qquad (2.4)$$

where $c_g = \sqrt{B_1^2 + gh}$.

Let us derive the expressions for Riemann invariants for the equations (2.3). To do this we find the eigenvectors of system (2.3). The left eigenvectors of (2.3) are following: $(c_g/h \quad 1 \quad 0 \quad 0)$, $(-c_g/h \quad 1 \quad 0 \quad 0)$, $(0 \quad 0 \quad 1 \quad 1)$, $(0 \quad 0 \quad 1 \quad -1)$. Multiplying equations (2.3) to the first eigenvector yields

$$\frac{\partial u_1}{\partial t} + \frac{c_g}{h}\frac{\partial h}{\partial t} + (u_1 + c_g)\left(\frac{\partial u_1}{\partial x} + \frac{c_g}{h}\frac{\partial h}{\partial x}\right) = -g\frac{\partial b}{\partial x} \qquad (2.5)$$

Introducing the function $\varphi(h) = \int \frac{c_g}{h} dh$ we rewrite the equation (2.5) in the following form:

$$\frac{\partial r}{\partial t}+\left(u_{1}+c_{g}\right)\frac{\partial r}{\partial x}=-g\frac{\partial b}{\partial x}, \text{ where } r=u_{1}+\varphi(h) \tag{2.6}$$

We note that the function $\varphi(h)$ cannot be expressed in elementary functions and is expressed in terms of elliptic integrals. However it is strictly increasing function. As the consequence the inverse function $\varphi^{-1}$ exists.

Multiplying (2.3) to other eigenvectors yields:

$$\frac{\partial s}{\partial t}+\left(u_{1}-c_{g}\right)\frac{\partial s}{\partial x}=-g\frac{\partial b}{\partial x} \text{ where } s=u_{1}-\varphi(h) \tag{2.7}$$

$$\frac{\partial p}{\partial t}+\left(u_{1}-B_{1}\right)\frac{\partial p}{\partial x}=0 \text{ where } p=u_{2}+B_{2} \tag{2.8}$$

$$\frac{\partial q}{\partial t}+\left(u_{1}+B_{1}\right)\frac{\partial q}{\partial x}=0 \text{ where } q=u_{2}-B_{2} \tag{2.9}$$

The functions $r$, $s$, $p$, $q$ are called the Riemann invariants and the system (2.4, 2.6-2.9) is called the shallow water magnetohydrodynamic equation system in the Riemann invariants form.

The expressions for the velocities $u_1, u_2$, fluid depth $h$ and magnetic field $B_2$ in terms of Riemann invariants are following:

$$u_{1}=\frac{r+s}{2}, \varphi(h)=\frac{r-s}{2}, u_{2}=p+q, B_{2}=p-q \tag{2.10}$$

According to the hyperbolic equations theory a Riemann wave is defined as the solution of (2.4, 2.6-2.9) in which all but one Riemann invariants remain constant. However, classical Riemann wave solutions do not satisfy equations (2.4, 2.6-2.9) due to the presence of function $-g\partial b/\partial x$ in the right-hand side of the equations.

We define the magnetogravity Riemann wave turned back as the solution satisfying the (2.6, 2.8-2.9) equations identically, and the magnetogravity Riemann wave turned forward as the solution satisfying the (2.7-2.9) equations identically. Similarly Alfwenic Riemann waves are defined as the solutions satisfying the (2.6-2.7, 2.9) or (2.6-2.8) equations identically. The reasons for these definitions will be seen below.

Let $p = p_0 = const$, $q = q_0 = const$ in some area of $(x,t)$, then equations (2.8-2.9) are identically satisfied in this area. Let us find the conditions when the expression for $r(x,t)$ satisfying equation (2.6) identically in the mentioned area exists. For that purpose we show the expression $u_1 + c_g$ is dependent on $s$ (maybe on $s$ and $r$) and $u_1 - c_g$ is dependent on $r$ (maybe $r$ and $s$). Definitely, $c_g$ is the function of $h$, and hence the function of $\varphi(h)$. Thus if $u_1 \pm c_g = f(u_1 \mp \varphi)$ then $c_g = -\varphi + const$ and it is not the case. Consequently, $u_1 + c_g$ is dependent on $s$ and $u_1 - c_g$ is dependent on $r$.

Functions $r(x,t)$ and $s(x,t)$ are linearly independent so the factor before the $(u_1 + c_g)$ has to be zero for the equation (2.6) to be satisfied identically, and thus $\partial r / \partial x \equiv 0$. However if $\partial r / \partial t \equiv 0$ too the equation (2.6) cannot be satisfied. Hence $r(x,t)$ is the function of time only, $r = r(t)$, therefore $\frac{\partial}{\partial x}\left(\frac{\partial r}{\partial t}\right) \equiv 0$ and $\frac{\partial}{\partial x}\left(-g \frac{\partial b}{\partial x}\right) \equiv 0$. So the solution satisfying the equation (2.6) can exist only for underlying surface $b(x)$ determined by $\frac{\partial^2}{\partial x^2}(b) \equiv 0$ equation, i.e. $b = kx + b_0$. The magnetogravity wave turned back does not exist for other underlying surfaces. It can be similarly shown the magnetogravity Riemann wave turned forward exists only for $\partial b / \partial x = k \equiv const$. Thus the simple wave solutions only exist for underlying surfaces that are slopes of constant inclination and furthermore we suppose $\partial b / \partial x = k \equiv const$.

## 3. Simple wave solutions for SMHD equations over a slope.

### 3.1. Continuous solutions

In this section we study simple wave solutions for shallow water magnetohydrodynamic equations over a slope which is particular case of initial equations (1.1-1.6) with $\partial b / \partial x = k \equiv const$. Let us consider magnetogravity Riemann wave turned back. In this case we have to satisfy equations (2.6, 2.8-2.9) identically. It was shown in previous section that $p = p_0 = const$, $q = q_0 = const$ satisfy equations (2.8-2.9), and it is easy to see that for $\partial b / \partial x = k \equiv const$ the expression

$$r = -gkt + r_0 \tag{3.1}$$

satisfies the equation (2.6) identically. We consider now the equation (2.7):

$$\frac{\partial s}{\partial t} + (u_1 - c_g)\frac{\partial s}{\partial x} = -gk \tag{3.2}$$

The equation (3.2) transforms along the characteristics

$$\frac{dx}{dt} = u_1 - c_g \tag{3.3}$$

to the following form:

$$\frac{\partial s}{\partial t} + \frac{dx}{dt}\frac{\partial s}{\partial x} = -gk \Leftrightarrow \frac{ds}{dt} = -gk \tag{3.4}$$

Integrating equation (3.4), we obtain

$$s(X(t),t) = \int_0^t \frac{ds}{dt} dt = -gkt + s(X(0),0) \tag{3.5}$$

Substituting expression (3.5) into equation (3.3) we get

$$\frac{dx}{dt} = -gkt + \frac{s(X(0),0) + r_0}{2} - c_g\bigg|_{\substack{r=-gkt+r_0 \\ s=-gkt+s(X(0),0)}} \quad (3.6)$$

The variable $c_g$ remains constant along the characteristics (3.3). Indeed,

$$\varphi(h) = \frac{r-s}{2} \text{ so } \varphi(h) = \frac{r_0 - s(X(0),0)}{2} = const \text{ along the}$$

characteristics (3.3). As $\varphi$ is bijective function then $h = const$ and $c_g = c_g(h) = const$ along the characteristics (6.3).

Integrating (3.6) we obtain the explicit expression for $X(t)$:

$$X(t) = \int_0^t \frac{dx}{dt} dt = -\tfrac{1}{2}gkt^2 +$$
$$+ \frac{s(X(0),0) + r_0}{2} t + c_g\big(h(X(0),0)\big)t + X(0) \quad (3.7)$$

The characteristics (3.7) are parabolas in the $(x,t)$ plane. It should be noted in case of flat surface the characteristics are straight lines:

$$X(t) = (u_1 + c_g)t + X(0) \quad (3.8)$$

and the change of variables

$$\begin{cases} \tilde{x} = x + \frac{1}{2}gkt \\ \tilde{t} = t \end{cases} \quad (3.9)$$

transforms parabolas (3.7) to straight lines (3.8). This change of variables will be used in the following to reduce the discontinuity decay problem on a slope to the same problem on flat surface.

For the magnetogravity Riemann wave turned forward the following relations hold true:

$$p = p_0, \; q = q_0, \; s = -gkt + s_0 \quad (3.10)$$
$$r(X(t),t) = -gkt + r(X(0),0) \quad (3.11)$$

$$X(t) = -\tfrac{1}{2}gkt^2 + \frac{r(X(0),0)+s_0}{2}t + c_g\left(h(X(0),t)\right)t + X(0) \qquad (3.12)$$

If $\dfrac{\partial s}{\partial x} > 0$ holds in some domain of $(x,t)$ plane for the Riemann magnetogravity wave turned back then the integral curves (3.3) are divergent. Taking into account expressions (3.1, 3.5) we get $u_1 = (r_0 + s)/2$, thus $\dfrac{\partial u}{\partial x} = \tfrac{1}{2}\dfrac{\partial s}{\partial x}$, and $\dfrac{\partial u}{\partial x} > 0$. Differentiating $r = u_1 + \varphi$ with respect to $x$, we yield $\dfrac{\partial u}{\partial x} + \dfrac{\partial \varphi}{\partial h}\dfrac{\partial h}{\partial x} = 0$, so $\dfrac{\partial h}{\partial x} < 0$ for $\dfrac{\partial \varphi}{\partial h} > 0$. Hence we have a magnetogravity rarefaction wave. If $\dfrac{\partial s}{\partial x} < 0$ in some domain $(x,t)$ then the integral curves are convergent, and we have a compression wave. In the domain $(x,t)$, in which $\dfrac{\partial s}{\partial x} = 0$, the characteristics are parallel lines, and we have a domain of uniformly accelerated flow.

The same results (except for sign) may be obtained for the Riemann magnetogravity wave turned forward ($s(x,t) = s_0 = const$). For $\dfrac{\partial r}{\partial x} < 0$ we have a rarefaction wave, for $\dfrac{\partial r}{\partial x} > 0$ we have a compression wave, for $\dfrac{\partial r}{\partial x} = 0$ we have a domain of uniformly accelerated flow.

Using the equations (3.1, 3.5-3.6, 3.10-3.12) and $p = const$, $q = const$, we get the relations for magnetogravity waves. For a magnetogravity Riemann wave turned back the following relations

$$B_1(x,t)h(x,t) = B_1(x_0,0)h(x_0,0)$$
$$B_2(x,t) = B_2(x_0,0)$$
$$u_2(x,t) = u_2(x_0,0)$$
$$u_1(x,t) + \varphi(x,t) + gkt = u_1(x_0,0) + \varphi(x_0,0)$$
(3.13)

are satisfied in the domain of the wave. Moreover, along the lines

$$\frac{dx}{dt} = u_1(x_0,0) - c_g(x_0,0) - gkt \qquad (3.14)$$

the equation

$$u_1(x,t) - \varphi(x,t) + gkt = u_1(x_0,0) - \varphi(x_0,0) \qquad (3.15)$$

is satisfied as well.

There $\varphi(x_0,0) = \varphi(h(x_0,0))$.

For a magnetogravity Riemann wave turned forward the following relations

$$B_1(x,t)h(x,t) = B_1(x_0,0)h(x_0,0)$$
$$B_2(x,t) = B_2(x_0,0)$$
$$u_2(x,t) = u_2(x_0,0)$$
$$u_1(x,t) - \varphi(x,t) + gkt = u_1(x_0,0) - \varphi(x_0,0)$$
(3.16)

are satisfied in the domain of the wave. Moreover, along the lines

$$\frac{dx}{dt} = u_1(x_0,0) + c_g(x_0,0) - gkt \qquad (3.17)$$

the equation

$$u_1(x,t) + \varphi(x,t) + gkt = u_1(x_0,0) + \varphi(x_0,0) \qquad (3.18)$$

is satisfied as well.

Let us consider an Alfvenic Riemann wave satisfying (2.6-2.8) equations. For this wave we obtain that the following relations

$$u_1(x,t) + gkt = u_1(x_0,0)$$
$$h(x,t) = h(x_0,0)$$
$$B_1(x,t) = B_1(x_0,0)$$
$$u_2(x,t) + B_2(x,t) = u_2(x_0,0) + B_2(x_0,0)$$
(3.19)

are satisfied in the domain of the wave. Hereafter, along the characteristics

$$\frac{dx}{dt} = u_1 + B_1 \qquad (3.20)$$

the equation

$$u_2(x,t) - B_2(x,t) = u_2(x_0,0) - B_2(x_0,0) \qquad (3.21)$$

is satisfied as well.

For the Alfvenic Riemann wave satisfying equations (2.6-2.7, 2.9) the following relations

$$u_1(x,t) + gkt = u_1(x_0,0)$$
$$h(x,t) = h(x_0,0)$$
$$B_1(x,t) = B_1(x_0,0)$$
$$u_2(x,t) - B_2(x,t) = u_2(x_0,0) - B_2(x_0,0)$$
(3.22)

are satisfied in the domain of the wave. Hereafter, along the characteristics

$$\frac{dx}{dt} = u_1 - B_1 \qquad (3.23)$$

the equation

$$u_2(x,t) + B_2(x,t) = u_2(x_0,0) + B_2(x_0,0) \qquad (3.24)$$

is satisfied as well.

We note that in Alfvenic Riemann waves all characteristics can be obtained from each other using a parallel translation. They are also parabolic curves same as for characteristics of magnetogravity waves.

We consider next the practically important special case of Riemann waves. The backward Riemann wave is termed centered wave if the characteristics (3.3) form a group of curves that come out of one point $(x_0, t_0)$. Let us denote the parameter assuming all values from the segment $\left[ \lim_{x \to x_0 - 0} (u_1(x,t_0) - c_g(x,t_0)), \lim_{x \to x_0 + 0} (u_1(x,t_0) - c_g(x,t_0)) \right]$ as $u'$.

Then the solution is determined by the equations

$$\begin{aligned} B_1(x,t)h(x,t) &= B_1(x_0,0)h(x_0,0) \\ B_2(x,t) &= B_2(x_0,0) \\ u_2(x,t) &= u_2(x_0,0) \\ u_1(x,t) - \varphi(x,t) + gkt &= u_1(x_0,0) - \varphi(x_0,0) \end{aligned} \qquad (3.25)$$

satisfied in the domain of the wave and the equation

$$u_1(x,t) + \varphi(x,t) + gkt = u_1(x_0,0) + \varphi(x_0,0) \qquad (3.26)$$

satisfied along the lines

$$\frac{dx}{dt} = u' - gkt \qquad (3.27)$$

As $\varphi - c_g$ is monotone function of $h$ then the $u_1(x_0,0)$ и $h(x_0,0)$ on each characteristics are uniquely determined by the relations $u_1(x_0,0) - \varphi(x_0,0) = const$ and $u' = u_1(x_0,0) - c_g(x_0,0)$. The equations (3.25-3.27) determine all the parameters in the centered magnetogravity wave turned back.

For a centered magnetogravity wave turned forward we denote the parameter assuming all values from the segment $\left[ \lim_{x \to x_0 - 0} (u_1(x,t_0) + c_g(x,t_0)), \lim_{x \to x_0 + 0} (u_1(x,t_0) + c_g(x,t_0)) \right]$

as $u'$. Then the solution is determined by the equations

$$\begin{aligned} B_1(x,t)h(x,t) &= B_1(x_0,0)h(x_0,0) \\ B_2(x,t) &= B_2(x_0,0) \\ u_2(x,t) &= u_2(x_0,0) \\ u_1(x,t) + \varphi(x,t) + gkt &= u_1(x_0,0) + \varphi(x_0,0) \end{aligned} \qquad (3.28)$$

satisfied in the domain of the wave and the equations

$$u_1(x,t) - \varphi(x,t) + gkt = u_1(x_0,0) - \varphi(x_0,0) \qquad (3.29)$$

satisfied along the lines

$$\frac{dx}{dt} = u' + gkt \qquad (3.30)$$

There $u_1(x_0,0)$ and $h(x_0,0)$ on each characteristics can be found from $u_1(x_0,0) + \varphi(x_0,0) = const$ and $u'$. The system (3.28-3.30) determines all the parameters in the centered magnetogravity wave turned forward.

## 3.2. Discontinuous solutions on slope. The jump conditions.

In this section the conditions that must be satisfied on the discontinuity lines are obtained. For this purpose we rewrite the equations (1.1-1.5) in the divergent form:

$$\begin{cases} \dfrac{\partial h}{\partial t} + \dfrac{\partial hu_1}{\partial x} = 0 \\ \dfrac{\partial hu_1}{\partial t} + \dfrac{\partial \left(hu_1^2 - hB_1^2 + 1/2gh^2\right)}{\partial x} = -g\dfrac{\partial b}{\partial x} \\ \dfrac{\partial hu_2}{\partial t} + \dfrac{\partial \left(hu_1 u_2 - hB_1 B_2\right)}{\partial x} = 0 \\ \dfrac{\partial hB_1}{\partial t} = 0 \\ \dfrac{\partial hB_2}{\partial t} + \dfrac{\partial \left(hu_1 B_2 - hB_1 u_2\right)}{\partial x} = 0 \end{cases} \qquad (3.31)$$

Integration of (3.31) on an arbitrary domain $G$ homeomorphic to square in the $(x,t)$ plane yields

$$\begin{cases} \iint_G \left( \dfrac{\partial h}{\partial t} + \dfrac{\partial hu_1}{\partial x} \right) dG = 0 \\[4pt] \iint_G \left( \dfrac{\partial hu_1}{\partial t} + \dfrac{\partial \left( hu_1^2 - hB_1^2 + 1/2\, gh^2 \right)}{\partial x} \right) dG = \iint_G \left( -g \dfrac{\partial b}{\partial x} \right) dG \\[4pt] \iint_G \left( \dfrac{\partial hu_2}{\partial t} + \dfrac{\partial \left( hu_1 u_2 - hB_1 B_2 \right)}{\partial x} \right) dG = 0 \\[4pt] \iint_G \left( \dfrac{\partial hB_1}{\partial t} \right) dG = 0 \\[4pt] \iint_G \left( \dfrac{\partial hB_2}{\partial t} + \dfrac{\partial \left( hu_1 B_2 - hB_1 u_2 \right)}{\partial x} \right) dG = 0 \end{cases} \quad (3.32)$$

Transforming volume integrals in (3.32) using the Green's formula we obtain

$$\oint_{\partial G} h\,dx - (hu_1)\,dt = 0$$

$$\oint_{\partial G} (hu_1)\,dx - (hu_1^2 - hB_1^2 + \tfrac{g}{2} h^2)\,dt = \oint_{\partial G} (gb)\,dt$$

$$\oint_{\partial G} (hu_2)\,dx - (hu_1 u_2 - hB_1 B_2)\,dt = 0 \qquad (3.33)$$

$$\oint_{\partial G} (hB_1)\,dx = 0$$

$$\oint_{\partial G} (hB_2)\,dx - (hu_1 B_2 - hB_1 u_2)\,dt = 0$$

Equations (3.33) represent the most general relations that are integral conservation laws and are valid for an arbitrary contour $\partial G$ and, in particular, for the contour including the discontinuity lines of an appropriate solution.

Let $x = x(t)$ be the equation of a jump-line and suppose it has a continuous tangent on the segment $[t_1, t_2]$. Assuming functions $u_1, u_2, B_1, B_2, h$ have a jump on the line $x = x(t)$ only and $b$ has no jump we denote

$$u_{1I}(t) = \lim_{x \to x(t)-0} u_1(x,t) \qquad u_{1II}(t) = \lim_{x \to x(t)+0} u_1(x,t)$$

$$u_{2I}(t) = \lim_{x \to x(t)-0} u_2(x,t) \qquad u_{2II}(t) = \lim_{x \to x(t)+0} u_2(x,t)$$

$$B_{1I}(t) = \lim_{x \to x(t)-0} B_1(x,t) \qquad B_{1II}(t) = \lim_{x \to x(t)+0} B_1(x,t)$$

$$B_{2I}(t) = \lim_{x \to x(t)-0} B_2(x,t) \qquad B_{2II}(t) = \lim_{x \to x(t)+0} B_2(x,t)$$

$$h_I(t) = \lim_{x \to x(t)-0} h(x,t) \qquad h_{II}(t) = \lim_{x \to x(t)+0} h(x,t)$$

(3.34)

Let $\partial G$ be the contour ABCE with lines AB and CE located infinitely close to the line of the jump $x(t)$ on the left and on the right sides respectively (fig. 1). Denoting the speed of discontinuity as $D = D(t) = x'(t)$, i.e. $dx = D(t)dt$, we obtain

$$\int_{AB}(Dh - hu_1)dt - \int_{CE}(Dh - hu_1)dt = 0$$

$$\int_{AB}(Dhu_1 - hu_1^2 + hB_1^2 - {g}/{2}\, h^2)dt - \int_{CE}(Dhu_1 - hu_1^2 + hB_1^2 - {g}/{2}\, h^2)dt = 0$$

$$\int_{AB}(Dhu_2 - hu_1u_2 + hB_1B_2)dt - \int_{CE}(Dhu_2 - hu_1u_2 + hB_1B_2)dt = 0 \qquad (3.35)$$

$$\int_{AB}(DhB_1)dt - \int_{CE}(DhB_1)dt = 0$$

$$\int_{AB}(DhB_2 - hu_1B_2 + hB_1u_2)dt - \int_{CE}(DhB_2 - hu_1B_2 + hB_1u_2)dt = 0$$

The contour ABCE is arbitrary, thus the equations (3.35) are equivalent to the following conditions for the integrands:

$$Dh_I - h_I u_{1I} = Dh_{II} - h_{II} u_{1II}$$

$$Dh_I u_{1I} - h_I u_{1I}^2 + h_I B_{1I}^2 - g/2\, h_I^2 = Dh_{II} u_{1II} - h_{II} u_{1II}^2 + h_{II} B_{1II}^2 - g/2\, h_{II}^2$$

$$Dh_I B_{1I} = Dh_{II} B_{1II} \qquad (3.36)$$

$$Dh_I u_{2I} - h_I u_{1I} u_{2I} + h_I B_{1I} B_{2I} = Dh_{II} u_{2II} - h_{II} u_{1II} u_{2II} + h_{II} B_{1II} B_{2II}$$

$$Dh_I B_{2I} - h_I u_{1I} B_{2I} + h_I u_{2I} B_{1I} = Dh_{II} B_{2II} - h_{II} u_{1II} B_{2II} + h_{II} u_{2II} B_{1II}$$

Let us consider the case $h_I \neq h_2$. Then the first three equations of (3.36) give

$$h_I B_I = h_{II} B_{II}$$

$$D = \frac{h_I u_{1I} - h_{II} u_{1II}}{h_I - h_{II}} \tag{3.37}$$

$$u_{1I} - u_{1II} = \pm(h_I - h_{II})\sqrt{\frac{g/2(h_I + h_{II}) + (B_{1I} h_I)^2/(h_I h_{II})}{h_I h_{II}}}$$

Substituting $D$ from the second relation of (3.37) to the last two equations of (3.36) and rearranging the terms we obtain

$$h_I h_{II}(u_{1I} - u_{1II})(u_{2I} - u_{2II}) = -(h_I - h_{II})(h_I B_{1I} B_{2I} - h_{II} B_{1II} B_{2II}) \tag{3.38}$$

$$h_I h_{II}(u_{1II} - u_{1I})(B_{2II} - B_{2I}) = (h_I - h_{II})(h_{II} u_{2II} B_{1II} - h_I u_{2I} B_{1I}) \tag{3.39}$$

If $B_{2I} = B_{2II}, u_{2I} = u_{2II}$ then (3.38, 3.39) are satisfied identically. In the other case we divide (3.38) to (3.39) and obtain $(u_{2I} - u_{2II})^2 = (B_{1I} - B_{1II})^2$ whence it follows that

$$u_{2I} - u_{2II} = \pm(B_{2I} - B_{2II}) \tag{3.40}$$

Substituting (3.40) to (3.38) and taking into account the third equation of (3.32) we obtain $h_I h_{II}(u_{1I} - u_{1II}) = \pm(h_I - h_{II}) h_I B_{1I}$ and thus

$$u_{1I} - u_{1II} = \pm(h_I - h_{II})\frac{h_I B_{1I}}{h_I h_{II}} \tag{3.41}$$

It is necessary for the sum of depths on both sides adjacent the discontinuity to be zero, $h_I + h_{II} = 0$, for the third equation of (3.37) and the equation (3.41) to be satisfied simultaneously. It can be only in the case when each depth equals to zero, i.e. the case of fluid absence. Thus the assumption that $B_2$ and $u_2$ have a discontinuity is incorrect if the equations (3.37) have a nontrivial solution.

Let us consider the other case when the free surface have no jump on the discontinuity, $h_I = h_{II}$. It follows from (3.37) the normal (to the discontinuity) components of velocity and magnetic fields have no jump as well, $B_{1I} = B_{1II}$, $u_{1I} = u_{1II}$. Thus the first three equations at the system (3.36) are satisfied identically. There are only two nontrivial relations on the discontinuity:

$$D = u_1 - B_1 \frac{B_{2I} - B_{2II}}{u_{2I} - u_{2II}}$$
$$\left(B_{2I} - B_{2II}\right)^2 = \left(u_{2I} - u_{2II}\right)^2$$
(3.42)

Rearranging (3.42) we get

$$D = u_1 \pm B_1$$
$$B_{2I} - B_{2II} = \mp\left(u_{2I} - u_{2II}\right)$$
(3.43)

The relations (3.43) are identically the ones obtained for the Alfvenic Riemann waves without discontinuity (3.16-3.18, 3.19-3.21).

Thus there are only two types of stable discontinuities with the nonzero mass flow through the discontinuity:
1) the discontinuity (3.37) with the free surface jump and transversal velocity and transversal magnetic field jumps, termed a magnetogravity shock wave;
2) the discontinuity (3.43) with the tangential velocity jump and the tangential magnetic field jump, termed an Alfvenic wave.

It should be noted that magnetogravity wave is the analog of a hydrodynamic jump for the classical shallow water equations, and the relations on this wave transform to ones on the hydrodynamic jump as $B_1 \to 0$. Magnetogravity shock wave is supersonic to the medium before the wave and subsonic to the medium after the wave as it is for classical hydrodynamic jump [19] in the shallow water theory [20].

In general, the system (3.36) accepts the third type of stable discontinuities with continuous tangential velocity component equal to the discontinuity velocity. It is termed the contact discontinuity. These discontinuities must be considered if the problem has different properties in the right and left half-spaces and these properties do no affect on the discontinuity decay solution. The example of such case is the fluid with different densities in the half-spaces separated by the discontinuity. Another example considered in present paper corresponds to the degeneration of an Alfvenic wave as $B_1 \to 0$. In this case the mass flow through the discontinuity equals zero and the tangential magnetic field and velocity field components act as the properties mentioned above.

## 3.3. The change of variables reducing the SMHD equations over a slope to ones over a flat plane

We consider the equation system (2.6-2.9). Taking into account $\dfrac{\partial b}{\partial x} = k = const$ we rearrange it in the form

$$\partial_t \begin{pmatrix} h \\ u_1 \\ u_2 \\ B_2 \end{pmatrix} + \begin{pmatrix} u_1 & h & 0 & 0 \\ c_g^2/h & u_1 & 0 & 0 \\ 0 & 0 & u_1 & -B_1 \\ 0 & 0 & -B_1 & u_1 \end{pmatrix} \partial_x \begin{pmatrix} h \\ u_1 \\ u_2 \\ B_2 \end{pmatrix} = \begin{pmatrix} 0 \\ -gk \\ 0 \\ 0 \end{pmatrix} \quad (3.44)$$

Using (3.9), we make the change of variables:

$$\begin{aligned} \tilde{x} &\to x + gkt^2/2 \\ \tilde{t} &\to t \end{aligned} \quad (3.45)$$

Thus

$$\begin{aligned} \dfrac{\partial}{\partial t} &= \dfrac{\partial}{\partial \tilde{t}} + gk\tilde{t}\dfrac{\partial}{\partial \tilde{x}} \\ \dfrac{\partial}{\partial x} &= \dfrac{\partial}{\partial \tilde{x}} \end{aligned} \quad (3.46)$$

This change of variables is nondegenerate, and after the following change of $u_1$

$$\tilde{u}_1 = u_1 + gkt \quad (3.47)$$

the system (3.44) transforms to

$$\partial_{\tilde{t}} \begin{pmatrix} h \\ \tilde{u}_1 \\ u_2 \\ B_2 \end{pmatrix} + \begin{pmatrix} \tilde{u}_1 & h & 0 & 0 \\ c_g^2/h & \tilde{u}_1 & 0 & 0 \\ 0 & 0 & \tilde{u}_1 & -B_1 \\ 0 & 0 & -B_1 & \tilde{u}_1 \end{pmatrix} \partial_{\tilde{x}} \begin{pmatrix} h \\ \tilde{u}_1 \\ u_2 \\ B_2 \end{pmatrix} = \begin{pmatrix} 0 \\ 0 \\ 0 \\ 0 \end{pmatrix} \quad (3.48)$$

After using the transformation (3.45, 3.46) the system (3.44) becomes the magnetohydrodynamic shallow water equation system on a flat surface (3.48) ($k=0$). This transformation is used later to solve the initial discontinuity decay problem over a slope reducing it to the discontinuity decay problem over a flat plane.

In this section we found particular wave solutions for the shallow water magnetohydrodynamic equations over a slope. Obtained solutions are used in the following section to find the exact explicit solutions of initial discontinuity decay problem over a slope.

## 4. Initial discontinuity decay problem for SMHD equations on a slope

Here we formulate the initial discontinuity decay problem for SMHD equations and list all possible wave configurations describing the nonlinear dynamics of initial discontinuity decay. We find the realization conditions for each wave configuration.

### 4.1. Initial discontinuity decay problem statement

We consider equations (1.1-1.5) with an arbitrary piecewise constant initial conditions for left $(x<0)$ and right $(x>0)$ half-spaces:

$$\begin{cases} t=0 \\ h=h_I, u_1=u_{1I}, u_2=u_{2I}, B_1=B_{1I}, B_2=B_{2I} \ for \ x<0 \\ h=h_{II}, u_1=u_{1II}, u_2=u_{2II}, B_1=B_{1II}, B_2=B_{2II} \ for \ x>0 \\ B_{1I}h_I = B_{1II}h_{II} \end{cases} \quad (4.1)$$

The discontinuity for two half-infinite magnetic fluids adjusted the $x=0$ plane at the initial time and satisfying (4.1) is called the initial discontinuity [21]. The determination of flow at $t>0$ for these initial conditions is called the initial discontinuity decay problem solution for SMHD equations.

Without the loss of generality it is assumed hereafter that the fluid depth in the right half-space is less than or equal to the fluid depth in the left half-space. It is shown below that in case of fluid absence in the right half-space the magnetic field component $B_1$ must be equal to zero in the left half-space, $B_{1I}=0$, and it leads to the absence of $B_1$ in the space-time domain of the solution. In this case solution is reduced to the classical dam break problem solution [19] with an additional convective transfer of tangential velocity and magnetic field components. It is assumed the right half-space magnetic fluid is at rest. The two above assumptions are easily satisfied when changing the coordinate system to one with proper axes direction and moving with prescribed velocity.

We note that in nontrivial case ($B_1 \neq 0$) the fluid depth is strictly positive in space-time region of solution because of magnetic field divergence-free condition. Thus if there is no fluid in one half-space then the normal magnetic field component in the other half-space degenerates. The case of fluid absence in the left and right half-spaces leads to the whole solution degeneration (all physical values are constant and equal zero) and is not considered. If $B_{1I}=B_{1II}=0$ then the problem reduces to the hydrodynamic initial discontinuity decay [19]. Indeed, in case of zero tangent magnetic field (an absence of magnetic field) equations (1.1-1.5) reduce to the classical shallow water system. In case of nonzero tangent magnetic field the solution degenerates. It is shown below two

Alfvenic waves merge and become the contact discontinuity, and tangent velocity and magnetic field components are transferred convectively. It corresponds to the classical dam break problem [19]. This case is not specially considered below.

**4.2 Discontinuity decay problem for a SMHD equations over a homogeneous surface.**

The initial discontinuity decay problem can be considered in the following way. Starting with the initial conditions (4.1) that represent some of the configurations mentioned below we change the parameters continuously, thus changing the solution. While changing parameters, we pass critical values which distinguish one configuration from another.

We consider the case of homogeneous surface, $k = 0$. The equation system (1.1-1.6) is invariant to the similarity transformation of $x$ and $t$: $t' = nt, x' = nx, n > 0$. Thus the solution provided below and the assumption of the solution uniqueness gives the self-similarity property of the solution. Conversely, the solution provided below and the assumption of the self-similarity property prove the existence and uniqueness of the solution.

Thus the determination of self-similar solutions of the initial discontinuity decay problem requires the matching of elementary solutions such as regions of constant flows, centered Riemann waves and hydrodynamic jumps, together with determination of their domains of applicability.

A self-similar picture of originating flow can be schematically shown in the space-time plane by one of five possible configurations (all rarefactions waves are centered ones):

'Two magnetogravity rarefaction waves, two Alfvenic waves' (fig. 2);

'Two magnetogravity shock waves, two Alfvenic waves' (fig. 3);

'Magnetogravity rarefaction wave turned back, magnetogravity shock wave, two Alfvenic waves' (fig. 4);

'Magnetogravity shock wave, magnetogravity rarefaction wave turned forward, two Alfvenic waves' (fig. 5);

'Two hydrodynamic rarefaction waves and a vacuum region between them' (fig. 6).

The latter case is the same as the corresponding configuration in the dam break problem [19] with additional convective transfer of tangent velocity and magnetic field components. This configuration is realized only when $B_1 \equiv 0$.

Let us show there are no more wave configurations. Indeed, the magnetogravity shock wave is supersonic with respect to the medium before the wave front and is subsonic with respect to the medium after the wave front. Thus two magnetogravity shock waves cannot propagate to the one half-space. Similarly, a magnetogravity rarefaction wave and a magnetogravity shock wave cannot propagate to the one half-space. Two magnetogravity rarefaction waves cannot propagate to one half-space as well because their outermost characteristics coincide, and they become one rarefaction wave. Moreover, the Alfvenic wave velocity is less than any magnetogravity wave velocity. It follows from the above conditions there can be no more than one magnetogravity wave (shock or rarefaction) and one Alfvenic wave propagating to the one half-space, so there are no more wave configurations that differ from those mentioned above.

### 4.2.1. Two magnetogravity rarefaction waves, two Alfvenic waves

Let the initial conditions determine the configuration 'two magnetogravity rarefaction waves, two Alfwen waves'. This configuration separates the flow domain to seven areas divided by six rays OA, OB, OC, OD, OE, OF (fig. 2). The OA ray determined by the equation $x = y_1 t$ separates the area *I* of constant flow where $u_1$, $u_2$, $B_1$, $B_2$ and $h$ remain constant and the area *III* of centered magnetogravity rarefaction wave turned back. According to (3.14), $y_1 = u_{1I} - c_{gI}$ as OA ray defines the characteristic of magnetogravity rarefaction wave turned back. OB ray defined by $x = y_2 t$ is also a characteristic separating the magnetogravity rarefaction wave turned back and the area *IV* of constant flow, so $y_2 = u_{1IV} - c_{gIV}$. OC and OD rays defined by $x = y_3 t$ and $x = y_4 t$ are corresponding to Alfvenic waves (3.43), so $y_3 = u_{1IV} - |B_{1IV}|$, $y_4 = u_{1V} + |B_{1V}|$. OE and OF rays corresponding to the magnetogravity rarefaction wave turned forward are defined by $x = y_5 t$ и $x = y_6 t$, where $y_5 = u_{1VI} + c_{VI}$, $y_6 = c_{gII}$.

We consider regions *VI*, *V* and *VI*. They are the regions of constant flow separated by the Alfvenic waves. It follows from (3.43) the values $u_2$ and $B_2$ in these regions are connected by the relations $u_{2IV} - B_{2IV} = u_{2V} - B_{2V}$, $u_{2V} + B_{2V} = u_{2VI} + B_{2VI}$ for $B_1 > 0$ and by $u_{2IV} + B_{2IV} = u_{2V} + B_{2V}$, $u_{2V} - B_{2V} = u_{2VI} - B_{2VI}$ for $B_1 < 0$. Thus

$$u_{2V} = \frac{1}{2}\left(u_{2VI} + u_{2IV} + \operatorname{sgn}(B_1)(B_{2VI} - B_{2IV})\right)$$
$$B_{2V} = \frac{1}{2}\left(\operatorname{sgn}(B_1)(u_{2VI} - u_{2IV}) + B_{2VI} + B_{2IV}\right)$$
(4.2)

As $u_2$ and $B_2$ remain constant in magnetogravity waves then $u_{2I} = u_{2III} = u_{2IV}$, $u_{2II} = u_{2VII} = u_{2VI}$, so

$$u_{2V} = \frac{1}{2}\left(u_{2II} + u_{2I} + \text{sgn}(B_1)(B_{2II} - B_{2I})\right)$$
$$B_{2V} = \frac{1}{2}\left(\text{sgn}(B_1)(u_{2II} - u_{2I}) + B_{2II} + B_{2I}\right)$$

(4.3)

Thus $u_2$ and $B_2$ are uniquely determined in the flow area at any $t > 0$.

As $u_1$, $B_1$ and $h$ remain constant on Alfwen waves, the following relations hold: $u_{1IV} = u_{1V} = u_{1VI}$, $B_{1VI} = B_{1V} = B_{1VI}$, $h_{IV} = h_V = h_{VI}$. Thus we denote the values of $u_1$, $B_1$ and $h$ in the regions *V* and *VI* as those with the subscript *IV*.

The region *III* is the magnetogravity rarefaction wave turned back (3.13-3.15), and the following equations holds in this region: $u_1 + \varphi(h) = const$, $B_1 h = const$. Thus $u_{1I} + \varphi(h_I) = u_{1IV} + \varphi(h_{IV})$, $B_{1I} h_I = B_{1IV} h_{IV}$. Similarly the region *VII* is the magnetogravity rarefaction wave turned forward (3.16-3.18), and the following equations hold in this region: $-\varphi(h_{II}) = u_{1IV} - \varphi(h_{IV})$, $B_{1II} h_{II} = B_{1IV} h_{IV}$. It immediately follows that

$$u_{1IV} = \frac{1}{2}\left(u_{1I} - \varphi(h_{II}) + \varphi(h_I)\right),$$

$$\varphi(h_{IV}) = \frac{1}{2}\left(u_I + \varphi(h_I) + \varphi(h_{II})\right).$$

The wave configuration considered is realized when $y_1 < y_2 < y_3 \leq y_4 < y_5 \leq y_6$. This expression is rewritten as:

$$u_{1I} - c_{gI} < u_{1IV} - c_{gIV} < u_{1IV} - |B_{1IV}| \leq u_{1IV} + |B_{1IV}| < u_{1IV} + c_{gIV} \leq c_{gII} \quad (4.4)$$

The last inequality of (4.4) transforms to the equality in the case of degeneration of the magnetogravity rarefaction wave turned forward. The equality $u_{1IV} - |B_{1IV}| = u_{1IV} + |B_{1IV}|$ holds only when there is no tangent magnetic field component. It corresponds to the case of merging of two Alfvenic waves to one. This wave is identically the contact discontinuity between two fluids with different normal magnetic field and velocity field components. Being only a particular case of two Alfvenic waves, this discontinuity will not be treated as a special case below.

The conditions $y_2 < y_3$ and $y_4 < y_5$ are satisfied with no additional assumptions because $c_g = \sqrt{B_1^2 + gh} > B_1$ holds in the flow region as the fluid depth is positive.

The only inequalities to be satisfied are $u_{1I} - c_{gI} < u_{1IV} - c_{gIV}$ and $u_{1IV} + c_{gIV} \leq c_{gII}$. To satisfy them it is necessary and sufficient to satisfy the inequality for fluid depths in regions *I, II, IV*:

$$h_{1I} \geq h_{1II} \geq h_{1IV} \tag{4.5}$$

Thus to find the necessary conditions for the considered configuration to be valid we must show the existence and uniqueness of $u_{1IV}$ and $h_{1IV}$ satisfying (4.5) and the invariant relations for the Riemann waves turned back and forward:

$$u_{1I} + \varphi(h_I) = u_{1IV} + \varphi(h_{IV}) \tag{4.6}$$

$$-\varphi(h_{II}) = u_{1IV} - \varphi(h_{IV}) \tag{4.7}$$

(4.6) and (4.7) together are equivalent to the following equation:

$$\varphi(h_{IV}) = \frac{1}{2}(u_{1I} + \varphi(h_I) + \varphi(h_{II})) \tag{4.8}$$

The function $\varphi(h) = \int \frac{1}{h}\sqrt{\frac{(B_1 h)^2}{h^2} + gh}$ is strictly increasing thus we can rewrite (5.2) using (4.8) as

$$u_I \leq \varphi(h_{II}) - \varphi(h_I) \tag{4.9}$$

When (4.9) is satisfied the configuration 'two magnetogravity rarefaction waves, two Alfvenic waves' is realized.

### 4.2.2. Magnetogravity rarefaction wave turned back, magnetogravity shock wave, two Alfvenic waves

This configuration separates the flow to six regions (fig. 4) divided by five rays OA, OB, OC, OD, OE. In the region *I* the flow parameters specified by the initial conditions (4.1) for the flow left half-space remain constant. Similarly, the flow parameters in the region *II* coincide with the initial conditions for the flow right half-space. OA and OB rays defined by $x = y_1 t$ and $x = y_2 t$, where $y_1 = u_{1I} - c_{gI}$, $y_2 = u_{1V} - c_{gV}$, bound the region *III* of the magnetogravity rarefaction wave turned back. OC and OD rays defined by $x = y_3 t$ and $x = y_4 t$, where $y_3 = u_{1V} - |B_{1V}|$, $y_4 = u_{1V} + |B_{1V}|$, are Alfvenic waves, and OE ray defined by $x = Dt$ is the magnetogravity shock wave.

The configuration realization conditions are:

$$u_{1I} - c_{gI} \leq u_{1V} - c_{gV} \leq u_{1V} - |B_{1V}| \leq u_{1V} + |B_{1V}| \leq D \tag{4.10}$$

Same as in the previous case the inequalities in (4.10) transform to equalities as the corresponding wave degenerates. The necessary and sufficient condition for the inequalities (4.10) to be hold is following:

$$h_{II} \leq h_{III} \leq h_{I} \tag{4.11}$$

The following relations hold in the region *IV*:

$$u_{2IV} = \frac{1}{2}\left(u_{2II} + u_{2I} + \text{sgn}(B_1)(B_{2II} - B_{2I})\right) \tag{4.12}$$

$$B_{2IV} = \frac{1}{2}\left(\text{sgn}(B_1)(u_{2II} - u_{2I}) + B_{2II} + B_{2I}\right) \tag{4.13}$$

Using the equalities $u_{1III} = u_{1IV} = u_{1V}$, $B_{1III} = B_{1IV} = B_{1V}$, $h_{III} = h_{IV} = h_{V}$ we can denote the values of $u_1$, $B_1$ and $h$ in regions *IV* and *V* as those with the subscript *III*.

In the region *III* the following condition on the magnetogravity rarefaction wave turned back $u_{1III}^{(1)} = u_{1I} + \varphi(h_I) - \varphi(h_{III})$ holds as well as the condition on the magnetogravity shock wave

$$u_{1III}^{(2)} = -(h_{II} - h_{III})\sqrt{\frac{g/2(h_{II} + h_{III}) + (B_{1I}h_I)^2/(h_{II}h_{III})}{h_{II}h_{III}}}.$$

The former function is the decreasing one in $h_{III}$ and the latter is increasing. Thus for the intersection point of these functions to exist and be unique on the $[h_{II}, h_I]$ segment it is necessary and sufficient for the inequalities $u_{1III}^{(1)}(h_{II}) \geq u_{1III}^{(2)}(h_{II})$ and $u_{1III}^{(1)}(h_I) \leq u_{1III}^{(2)}(h_I)$ to be satisfied. These inequalities are rewritten as:

$$u_{1I} \geq \varphi(h_{II}) - \varphi(h_I) \tag{4.14}$$

$$u_{1I} \leq (h_I - h_{II})\sqrt{\frac{g/2(h_I + h_{II}) + (B_{1I}h_I)^2/(h_I h_{II})}{h_I h_{II}}} \tag{4.15}$$

Thus the given configuration is realized when the conditions (4.14, 4.15) are satisfied.

It should be noted the particular case of equality in (4.14) corresponds to the magnetogravity shock wave degeneration.

### 4.2.3. Two magnetogravity shock waves, two Alfvenic waves

This configuration separates the flow area to five regions (fig. 3) with constant flow in each of them, divided by four rays OA, OB, OC, OD. In the region *I* the flow parameters specified by the initial conditions (4.1) for the flow left half-space remain constant. Similarly, the flow parameters in the region *II* coincide with the initial conditions for the flow right half-space. The OA ray defined by $x = D_1 t$ is the magnetogravity shock wave with the (3.37) conditions holding on it. Similarly, the OD ray defined by $x = D_2 t$ is the magnetogravity shock wave with the (3.37) conditions on it. OB and OC rays defined by $x = y_1 t$ and $x = y_2 t$ correspondingly, where $y_1 = u_{1III} - |B_{1III}|$, $y_2 = u_{1IV} + |B_{1IV}|$, are Alfvenic waves.

The necessary condition for this configuration to be realized is the following inequality:

$$D_1 \leq u_{1III} - |B_{1III}| \leq u_{1IV} + |B_{1IV}| < D_2 \tag{4.16}$$

The first and the last inequalities follow immediately from the expression for the magnetogravity shock waves velocity.

Considering regions *III, IV, V* we obtain the following expressions:

$$u_{2V} = \frac{1}{2}\left(u_{2VI} + u_{2IV} + \text{sgn}(B_1)(B_{2VI} - B_{2IV})\right) \quad (4.17)$$

$$B_{2V} = \frac{1}{2}\left(\text{sgn}(B_1)(u_{2VI} - u_{2IV}) + B_{2VI} + B_{2IV}\right) \quad (4.18)$$

Using the equalities $u_{1III} = u_{1IV} = u_{1V}$, $B_{1III} = B_{1IV} = B_{1V}$, $h_{III} = h_{IV} = h_V$ we can denote the values of $u_1$, $B_1$ and $h$ in regions *IV* and *V* as those with the subscript *III*.

It follows from the conditions on a shock waves that

$$u_{1III}^{(1)} = u_{1I} + (h_I - h_{III})\sqrt{\frac{g/2(h_I + h_{III}) + (B_{1I}h_I)^2/(h_I h_{III})}{h_I h_{III}}} \quad (4.19)$$

$$u_{1III}^{(2)} = -(h_{II} - h_{III})\sqrt{\frac{g/2(h_{II} + h_{III}) + (B_{1I}h_I)^2/(h_{II} h_{III})}{h_{II} h_{III}}} \quad (4.20)$$

Using the inequality for the shock waves $h_{III} \geq h_I \geq h_{II}$ we can see that the first expression is the decreasing function of $h_{III}$ and the second is the increasing one. Thus for these functions to have the unique intersection on $[h_I, +\infty)$ half-interval it is necessary and sufficient that $u_{1III}^{(1)}(h_I) \geq u_{1III}^{(2)}(h_I)$. Thus we get the following condition on the initial data:

$$u_{1I} \geq (h_I - h_{II})\sqrt{\frac{g/2(h_I + h_{II}) + (B_{1I}h_I)^2/(h_I h_{II})}{h_I h_{II}}} \quad (4.21)$$

When the inequality (4.21) holds the configuration 'two magnetogravity shock waves, two Alfvenic waves' is realized.

It should be noted the case of equality in (4.21) corresponds to the left magnetogravity shock wave degeneration.

**4.2.4. Two hydrodynamic rarefaction waves and a vacuum region between them**

In case of $B_{1I} = B_{1II} = 0$ magnetogravity rarefaction waves transform to a hydrodynamic rarefaction waves and the configuration 'two hydrodynamic rarefaction waves and a vacuum region between them' can be realized. The condition of its realization is $u_{1I} < -2c_{gI} - 2c_{gII}$ [19].

**4.2.5. Initial discontinuity decay problem solution**

The initial discontinuity decay problem for the magnetohydrodynamic shallow water equations over a flat plane is solved for an arbitrary initial data satisfying the following conditions

$$\left[ \begin{array}{l} \begin{cases} B_{1I} h_I = B_{1II} h_{II} \\ h_{II} \neq 0 \\ h_I \geq h_{II} > 0 \\ u_{1II} = 0 \end{cases} \\ \begin{cases} h_{II} = 0 \\ B_{1I} = B_{1II} = 0 \\ u_{1II} = 0 \end{cases} \end{array} \right. \quad (4.22)$$

When

$$u_{1I} \geq (h_I - h_{II})\sqrt{\frac{g/2(h_I + h_{II}) + (B_{1I}h_I)^2/(h_I h_{II})}{h_I h_{II}}}$$

the configuration 'two magnetogravity shock waves, two Alfvenic waves' is realized.

When

$$\begin{cases} u_{1I} > \varphi(h_{II}) - \varphi(h_I) \\ u_{1I} < (h_I - h_{II})\sqrt{\dfrac{g/2(h_I + h_{II}) + (B_{1I}h_I)^2/(h_I h_{II})}{h_I h_{II}}} \end{cases}$$

the configuration 'magnetogravity rarefaction wave turned back, magnetogravity shock wave, two Alfvenic waves' is realized.

When

$$u_{1I} \leq \varphi(h_{II}) - \varphi(h_I)$$

the configuration 'two magnetogravity rarefaction waves, two Alfvenic waves' is realized.

When

$$\begin{cases} B_{1I} = B_{1II} = 0 \\ u_{1I} < -2c_{gI} - 2c_{gII} \end{cases}$$

the configuration 'two hydrodynamic rarefaction waves and a vacuum region between them' is realized.

The Riemann problem solution found before forms a basis for development of finite-volume numerical methods to compute continuous and discontinuous solutions without capturing of discontinuities [22-24].

## 4.3. The reverse change of variables and solution of the initial discontinuity decay problem for SMHD equations over a slope

It is shown before the type of wave configuration is determined by the initial conditions (4.1) only. The initial conditions are invariant to the transformation (7.2, 7.4), and the conditions for the type of a wave configuration on a slope are the same as the conditions for the flat plane. Thus to find the solution of an initial discontinuity decay problem on a slope we use the initial discontinuity decay problem solution on a flat plate obtained before and perform the change of variables reverse to (7.2, 7.4):

$$\begin{cases} x = \tilde{x} - \frac{1}{2} gkt^2 \\ t = \tilde{t} \\ u = \tilde{u} - gkt \end{cases} \quad (4.23)$$

It should be noted the structure of the Riemann problem solution on a slope is the same as that on a flat plane. Indeed, the characteristics in case of slope are parabolas whereas the characteristics in case of flat plane are straight lines (fig. 2-6, 7-10). These characteristics are tangent in the initial point.

## Conclusion

In present article the nonlinear dynamics of shallow water magnetohydrodynamic flows of heavy fluid is studied. It is shown simple wave solutions exist only when the surface is slope. All simple wave solutions for slopes are found: Alfvenic waves and magnetogravity waves. The change of variables transforming SMHD equations over a slope to ones over a flat plane is found. The exact explicit solution of initial discontinuity decay problem over a flat plane and over a slope is found. It is shown these solutions are represented by one of the following configurations: 'Two magnetogravity rarefaction waves, two Alfvenic waves', 'Two magnetogravity shock waves, two Alfvenic waves', 'Rarefaction magnetogravity wave turned back, magnetogravity shock wave, two Alfvenic waves', 'Magnetogravity shock wave, magnetogravity rarefaction wave turned forward, two Alfvenic waves', 'Two hydrodynamic rarefaction waves and a vacuum region between them'. Despite of the same wave configurations in case of slope and flat plane, these solutions are drastically differ from each other. In case of flat plane the characteristics of waves are straight lines and in case of slope they are parabolas. The regions of constant flow in the flat plane solutions are transformed to the regions of constantly accelerated flow in case of slope.

It follows from the obtained results that the solution of initial discontinuity decay is the superposition of two solutions: the initial discontinuity decay solution for shallow water without magnetic field (with modified sound velocity $c_g = \sqrt{B_1^2 + gh}$) and two Alfvenic waves. When $B_1 \equiv 0$ two Alfvenic waves merge and become the contact discontinuity. The 'two hydrodynamic rarefaction waves and a vacuum region between them' configuration differs from other configurations and can be realized only when normal component of magnetic field equals to zero.

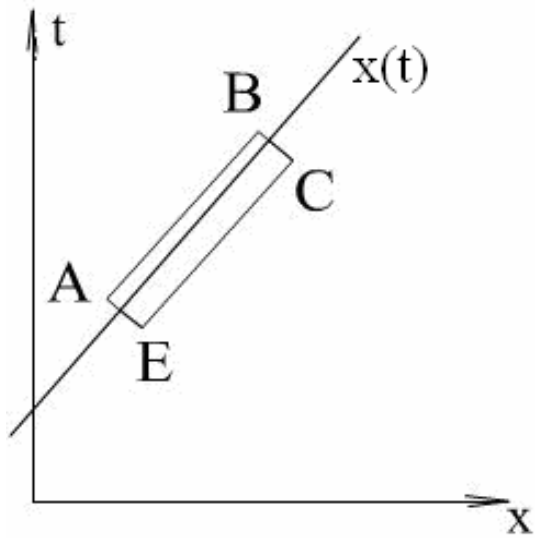

**Fig. 1** Contour ABCE with lines AB and CE located infinitely close to the line of the jump $x(t)$

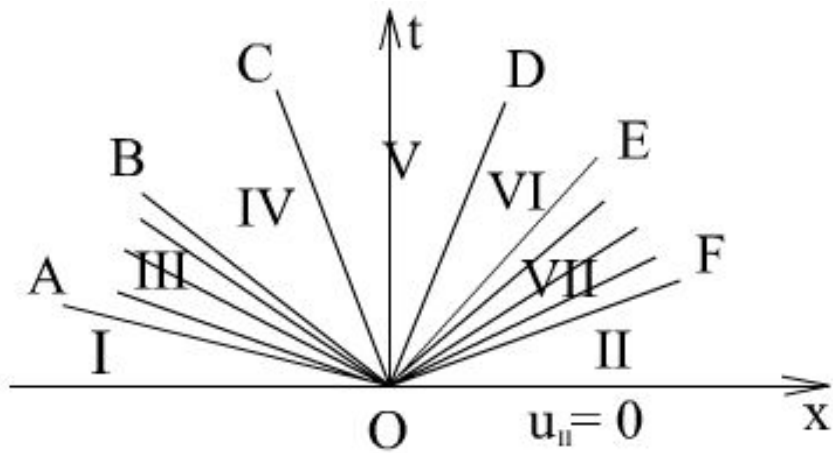

**Fig. 2** Two magnetogravity rarefaction waves, two Alfvenic waves
I, II, IV, V, VI – regions of constant flow; III, VII – magnetogravity rarefaction waves; OC, OD – Alfvenic waves.

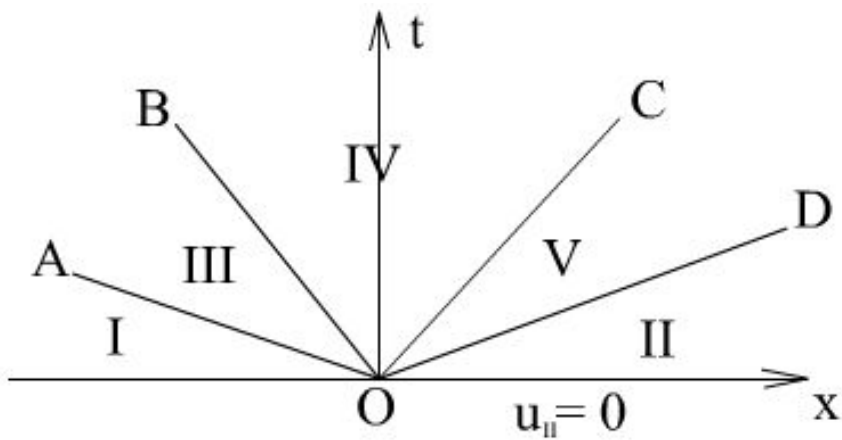

**Fig. 3** Two magnetogravity shock waves, two Alfvenic waves

I, II, III, IV, V – regions of constant flow; OA, OD – magnetogravity shock waves; OB, OC – Alfvenic waves.

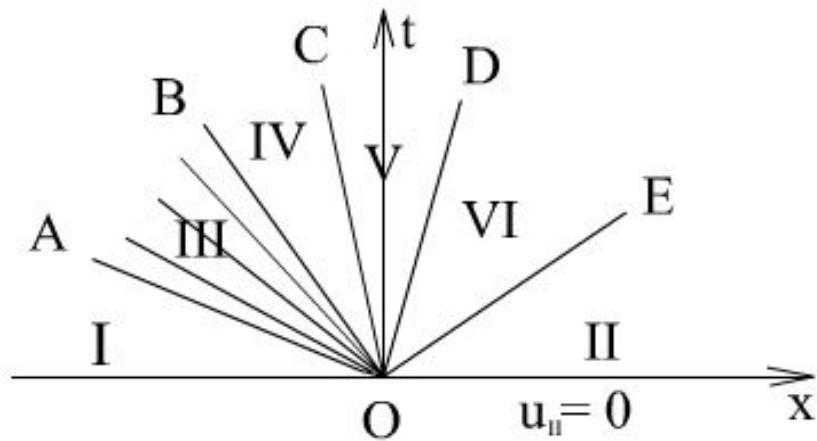

**Fig. 4** Magnetogravity rarefaction wave turned back, magnetogravity shock wave, two Alfvenic waves

I, II, IV, V, VI – regions of constant flow; III – magnetogravity rarefaction wave; OC, OD – Alfvenic waves. OE – magnetogravity shock wave

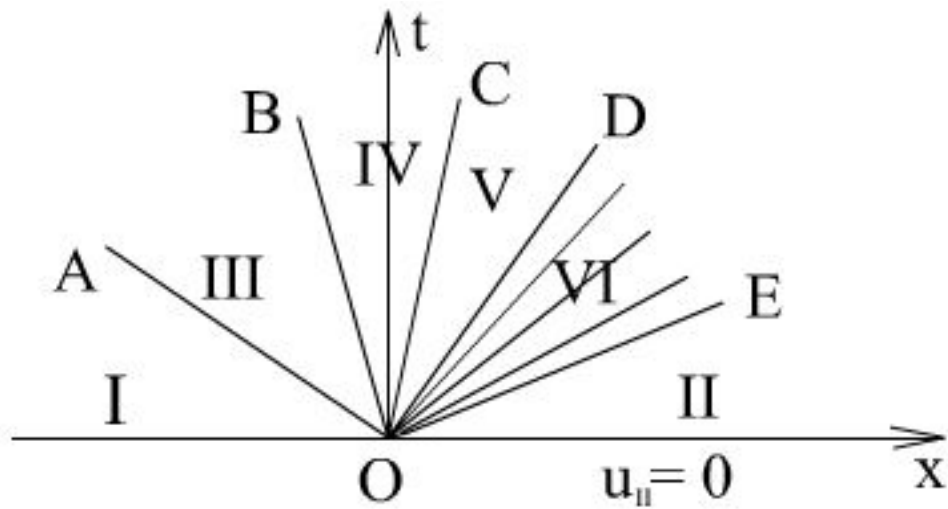

**Fig. 5** Magnetogravity shock wave, magnetogravity rarefaction wave turned forward, two Alfvenic waves

I, II, III, IV, V – regions of constant flow; IV – magnetogravity rarefaction wave; OA - magnetogravity shock wave; OB, OC - Alfvenic waves

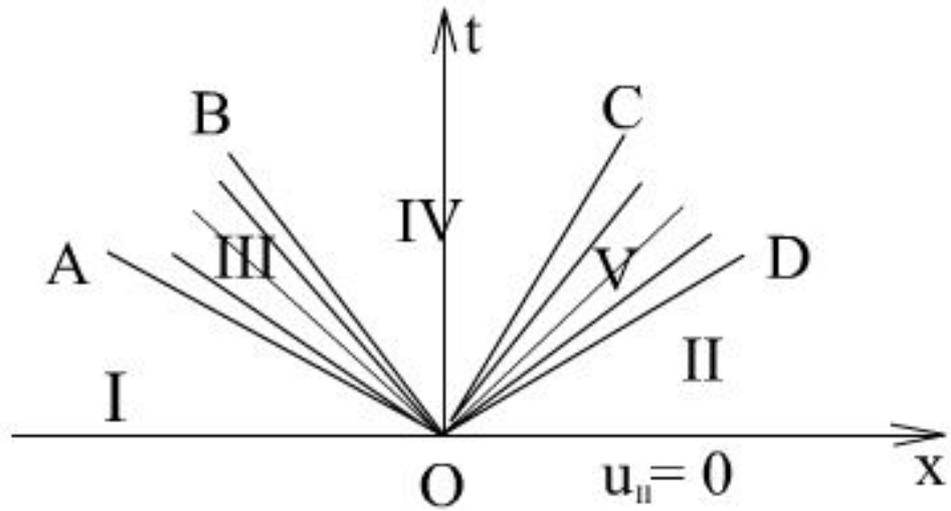

**Fig. 6** Two hydrodynamic rarefaction waves and a vacuum region between them

I, II - regions of constant flow; III, V - magnetogravity rarefaction waves; IV – vacuum region

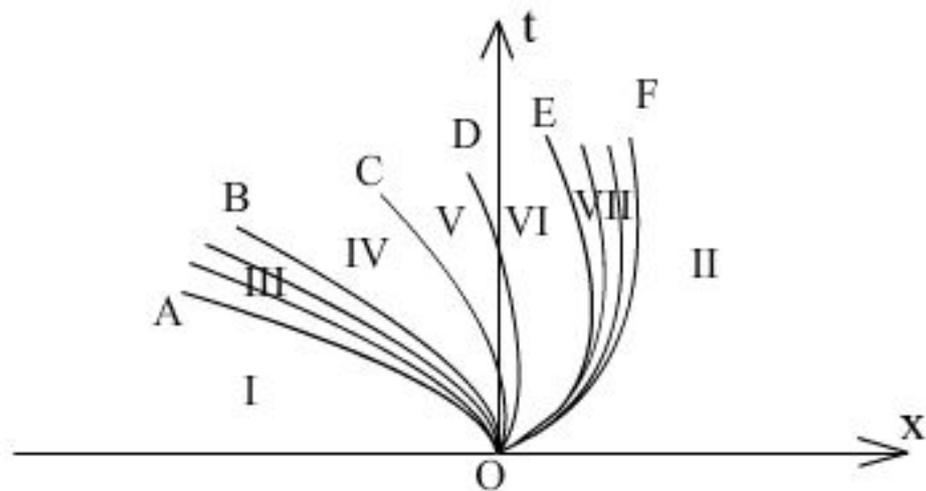

**Fig. 7** Two magnetogravity rarefaction waves, two Alfvenic waves over a slope

I, II, IV, V, IV – regions of uniformly accelerated flow; III, VII – magnetogravity rarefaction waves; OC, OD – Alfvenic waves

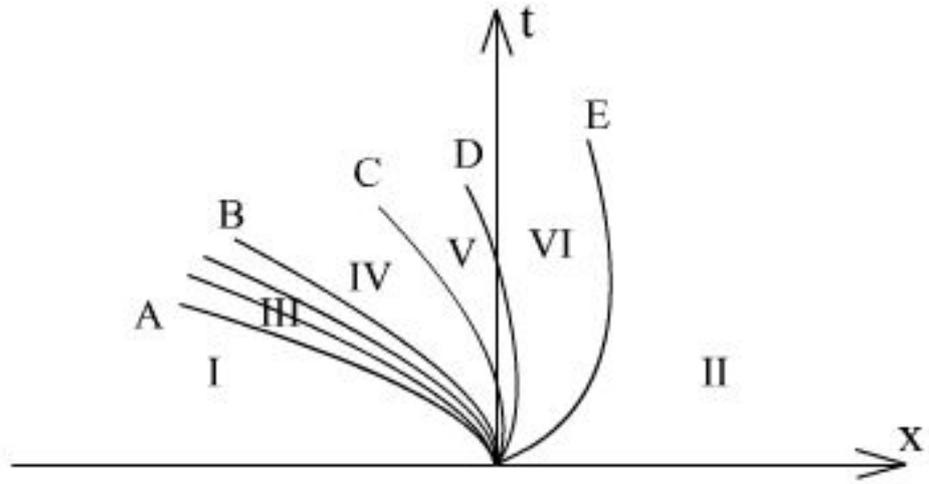

**Fig. 8** Magnetogravity wave turned back, magnetogravity shock wave, two Alfvenic waves over a slope

I, II, IV, V, IV – regions of uniformly accelerated flow; III – magnetogravity rarefaction wave; OC, OD – Alfvenic waves; OE – magnetogravity shock wave

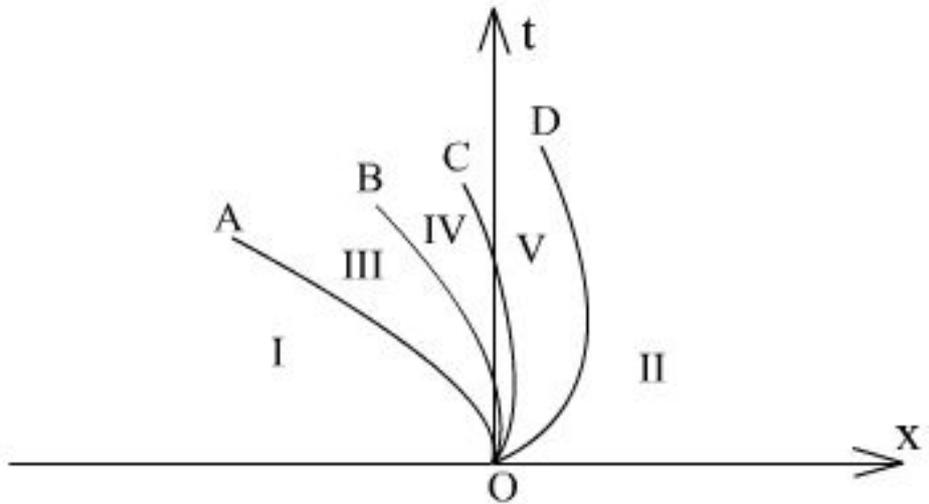

**Fig. 9** Two magnetogravity shock waves, two Alfvenic waves over a slope

I, II, III, IV, V – regions of uniformly accelerated flow; OA, OD – magnetogravity shock waves; OB, OC – Alfvenic waves.

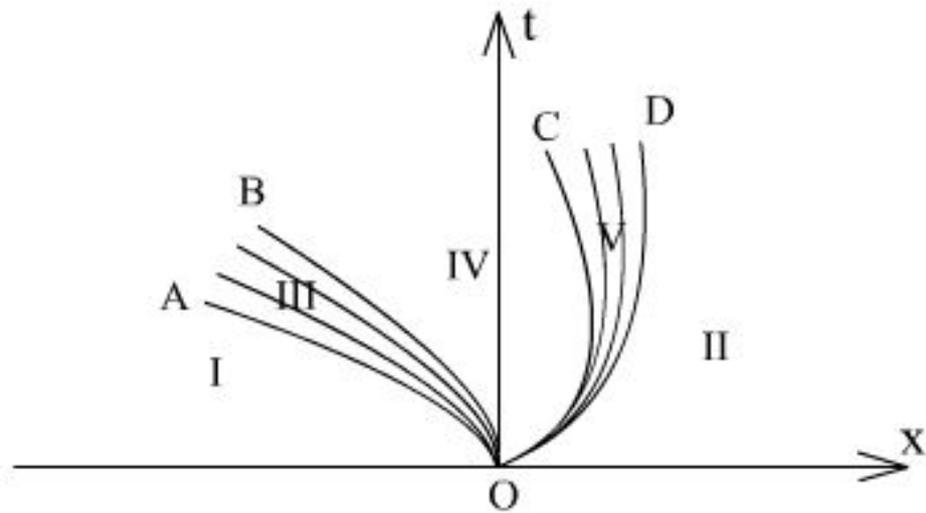

**Fig. 10** Two hydrodynamic rarefaction waves and a vacuum region between them over a slope

I, II - regions of uniformly accelerated flow; III, V - magnetogravity rarefaction waves; IV – vacuum region